\documentclass[12pt,thmsa]{article}
\usepackage{sw20lart}

\input tcilatex
\begin{document}

\title{A schematic age-structured compartment model of the impact of antiretroviral
therapy on HIV incidence and prevalence}
\author{L. F. Lopez$^{1}$ \and F. A. B. Coutinho$^{1}$ \and M. N. Burattini$^{1}$
\and E. Massad$^{1,2}$ \\
$^{(1)}${School of Medicine of the University of S\~{a}o Paulo }\\
{and LIM01/HCFMUSP, }\\
{Av. Dr. Arnaldo, 455, CEP 01246-903, SP, Brazil} \\
$^{(2)}${Department of Infectious and Tropical Diseases (Hon. Prof.), }\\
{London School of Hygiene and Tropical Medicine, }\\
{London, UK}}
\date{04 January 2005}
\maketitle

\begin{abstract}
A simple deterministic model is proposed to represent the basic aspects
concerning the effects of different antiretroviral treatment schedulles on
HIV\ incidence and prevalence of affected populations. The model mimics
current treatment guidelines applied in Brazil. However, the model does not
intend to fit the data with any acceptable degree of accuracy since
uncertainties on the values of the parameters and on the precise effect of
the treatment put some limits on the practical implications of our model
from which only orders of magnitude and some qualitative effects can be
deduced. So, this paper intends to provide a conceptual and mechanistic
understanding of the possible long term effects of treatment on the dynamics
of HIV transmission.

According to the model, the effect of the treatment depends on the level of
sexual activity of the subpopulations considered, being more pronounced on
the subpopulations with the highest sexual activity levels. Also,
inefficient treatment can be prejudicial depending on the level of sexual
activity and on the capacity to provide adequate treatment coverages to the
population affected.
\end{abstract}

\section{Introduction}

The natural history of the Human Immunodeficiency Virus (HIV) infection and
transmission is well understood today. It is now accepted that the viral
concentration in the circulating blood (viraemia) and other organic fluids
determines the probability of transmission \cite{mellors}. It has been
demonstrated by several authors that after infection the viraemia rapidly
increases in the first week to a high level, staying in that high levels for
some weeks, dropping afterwards to very low levels and then increasing
slowly for several years. After a period of 10 - 15 years, the increasing
level of viraemia reaches a level in which clinical manifestations begin,
coinciding with the dropping in the CD4 lymphocytes (the target cell of HIV)
counting and the development of the conditions that define the state of
full-blown AIDS.

So, it should be expected a large variation in the probability of
transmission of HIV along the natural course of the infection \cite{erb}.
The relative contribution of each of these distinct phases of viraemia seen
along the natural history of the infection to HIV transmission have only
recently been demonstrated \cite{nos2001}. This relative contribution may
have important consequences to the epidemiological pattern of HIV
transmission, as well as to the assessment of the impact of HIV treatment on
the epidemiology of HIV/AIDS. In addition, it should be expected that
differences in the probability of transmission in each of the distinct
phases of the infection may also have important consequences on the
evolution of HIV virulence \cite{nos}.

The use of the combined antiretrovirus therapies (ARVTs), particularly those
known as HAART (highly active antiretrovirus therapies) has demonstrated
excellent results in all the countries where people infected by HIV has
access to the treatment \cite{palela}. The results reported significant
reduction in deaths rates caused by opportunistic infections. For example,
in the U.S.A. there has been a reduction of 61\% between 1995 and 1997 in
the mortality due to AIDS. In Brazil, whose experience has been acclaimed
worldwide as one of the most successful attempts to control AIDS, the
government decided to provide ARVT to all HIV seropositive individuals who
fulfilled the treatment criteria. From 1996 onwards HAART has been used as a
standard treatment, with 120,000 patients being treated in 2002.

The treatment against HIV evolved rather rapidly. Just 4 years after the
identification of HIV as the causative agent of the new syndrome, AIDS,
zidovudine (AZT) the first drug for its treatment was licenced by the FDA.
In the next eight years three other nucleoside analogs (the same class as
zidovudine) were introduced. Concurrently, a better understanding of the
dynamics of HIV replication and drug resistance mechanisms caused a shift
from single- to combination-drug therapy. From 1995 to 1998 eight new
antiretroviral agents were approved, including protease inhibitors and
non-nucleosides transcriptase inhibitors . These latter drugs also provided
clinicians with a highly effective antiretroviral therapy, known as HAART
that reduces viral load by a factor of $10^{3}$. However, these new drugs
cause several important side effects that limit their use. Because of this,
current clinical guidelines propose a delaying in starting HIV treatment to
a point in which either the viral load is high (above 30000 copies per ml)
or the CD4 cell count is low. However, with the progression of the
infection, even in the persistence of treatment, HIV\ viral load tends to
increase again leading to a need to change the drugs in use or to
progression to AIDS \cite{mandelT}.

The discussion on the implications of antiretroviral antiretroviral
treatment has been restricted to clinical and virological aspects. However,
considering that different antiretroviral treatment strategies have distinct
effects on viral load, one crucial aspect to be considered is the
epidemiological consequences of a given antiretroviral treatment strategy on
the incidence of new HIV infections. On one hand, effective antiretroviral
treatment reduces viral load. On the other hand, it prolongs the
asymptomatic phase, probably the most important for transmission \cite
{nos2001} in the absence of antiretroviral treatment. Both effects are
intrinsically related to transmission, although in opposite directions.
Therefore, the critical question related to the choice of the best
antiretroviral strategy should take into account the epidemiological
consequences of different antiretroviral treatments. In spite of some
attempts to understand this crucial aspect, the best antiretroviral
treatment strategy\ is still to be defined \cite{levin},\cite{anderson1991}, 
\cite{blower}.

In this paper we propose a schematic mathematical model to analyze the
impact of current antiretroviral therapy on the incidence and prevalence of
HIV infection. In section 2, we point out some historical aspects of
Brazilian public health strategies related to HIV/AIDS treatment and its
impact on the epidemics in Brazil. Our model was constructed incorporating
the treatment concepts of the Brazilian program. However, due to
uncertainties in parameters, practical conclusions should be taken with
caution. In section 3, we derive an integral equation which allows the
calculation of the impact on HIV incidence of an antiretroviral treatment
program. This equation depends on a number of functions and parameters whose
forms and values are described in section 4. In section 5, we present
numerical results of several hypothetical scenarios and in section 6 we
discuss the model\'{}s limitations and draw some tentative conclusions
regarding possible epidemiological implications of our results.

\section{Motivation: the Brazilian experience with HAART}

Since 1991 the Brazilian government decided to provide ARVT to all HIV
seropositive individuals who fulfilled the treatment criteria \cite{MS}.
From 1996 onwards HAART has been used as a standard treatment, with 120,000
patients being treated in 2002. Its impact was immediately noted by health
authorities, in particular its dramatic reduction on AIDS\ mortality, and
this public health experience has been acclaimed worldwide as one of the
most successful attempts to control AIDS.

Figure 1 shows the reduction in the mortality by AIDS in the period between
1996 and 2001. The total number of averted deaths summed up to 90,000
patients in this period.

\begin{center}
\textbf{Figure 1}
\end{center}

In addition, it was observed an increase in the survival period of patients
with AIDS, from 5 to 58 months, a 12 fold increase, after the introduction
of HAART treatment. Also, there was a six-fold reduction in the number of
AIDS hospitalizations and a reduction of 54\% in the cost of treatment.

With respect to the impact of HAART on the incidence and prevalence of HIV
infection, only rough estimates are available. For instance, in 1992 the
World Bank projected the number of expected HIV infected individuals in
Brazil by 2002 as 1.2 million. However, the current estimates for 2002 is
around 600 thousands. It is difficult to attribute this reduction solely to
HAART, since other preventive practices have been highly stimulated by the
Brazilian government. For instance, the observed reduction in HIV infection
prevalence among sex workers from 18\% in 1996 to 6.1\% in 2000, and among
homosexuals from 10.8\% in 1999 to 4.7\% in 2001, has been attributed by
health authorities to the association of condom distribution and other
preventive measures adopted.

The most important indication of the possible impact of HAART on HIV
transmission in Brazil is the marked reduction in incidence rates of AIDS,
as shown in figure 2. It is noticeable that after a historical increase in
the incidences from 8 cases per 100,000 people per year in 1991 up to the
peak of 18 cases per 100,000 people per year attained in 1998, the number of
new cases dropped to 5 cases per 100,000 people per year in 2003. If we
consider that the universal treatment started in 1996, we may conclude that
after a delay of two years, the effect of HAART on the incidence of AIDS
could be observed.

\begin{center}
\textbf{Figure 2}
\end{center}

Notwithstanding, the observed reduction in HIV incidence and prevalence
cannot be entirely explained by the effects due to HAART. Additionally,
behavioral changes attributable to the preventive campaigns carried out in
Brazil simultaneously to the beginning of HAART may also have had a positive
impact on HIV transmission. For instance, there have been a marked increase
in regular condom use verified between 1999 and 2000 (42\% to 64\%).

The Brazilian experience motivated us to model the possible role of ARVT on
the incidence and prevalence of HIV. However, lack of data and uncertainties
on the precise effect of the treatment put some limits on the practical
implications of our model from which only orders of magnitude and some
qualitative effects can be deduced.

\section{The model: formalism}

\subsection{Sexually transmitted HIV}

As in previous papers, we consider a community of N individuals in steady
state with respect to time \cite{nos2001}, divided into classes according to
the contact pattern and transmission intensity of HIV \cite{nos1999}. For
simplicity, in this paper we consider that the interaction between
individuals from different classes is so rare that can be neglected. The
classes will be described in section 5. As the interclass interactions are
neglected, we describe a general formalism which applies to any and all
classes.

Let $N(a)da$ be the number of individuals with age between $a$ and $a+da$.
Among those, let $X(a)da$ be the number of susceptibles with age between $a$
and $a+da$. The number of unprotected sexual contacts per unit time those
individuals with age between $a$ and $a+da$ make with all other individuals
with ages between $a^{\prime }$ and $a^{\prime }+da^{\prime }$ is: 
\begin{equation}
\beta _{0}(a,a^{\prime })da^{\prime }X(a)da  \label{0}
\end{equation}

Let $Y_{1}(a^{\prime },\tau )da^{\prime }d\tau $ be the number of
individuals with age between $a^{\prime }$ and $a^{\prime }+da^{\prime }$
infected when aged between $\tau $ and $\tau +d\tau .$ The number of
unprotected sexual contacts per unit time those individuals with age between 
$a$ and $a+da$ make with all other individuals with ages between $a^{\prime
} $ and $a^{\prime }+da^{\prime }$ is 
\begin{equation}
\beta _{0}(a,a^{\prime })da^{\prime }\frac{Y_{1}(a^{\prime },\tau
)da^{\prime }d\tau }{N(a^{\prime })da^{\prime }}X(a)da  \label{0a}
\end{equation}

Let $g(a^{\prime }-\tau )$ be the probability of a susceptible individual to
get infected when making an unprotected sexual contact with an individual $%
Y_{1}(a^{\prime },\tau )da^{\prime }d\tau $. This depends on viral load
which in turn depends on the time since infection $(a^{\prime }-\tau )$, and
its form is given by equation (\ref{18a}) Therefore, the number of new
infections per unit time, due to individuals $Y_{1}(a^{\prime },\tau
)da^{\prime }d\tau $, is given by 
\begin{equation}
\beta _{0}(a,a^{\prime })g(a^{\prime }-\tau )\frac{Y_{1}(a^{\prime },\tau )}{%
N(a^{\prime })}da^{\prime }d\tau X(a)da  \label{0b}
\end{equation}

Let $Y_{2}(a^{\prime },\tau ,l)da^{\prime }d\tau dl$ be the number of
individuals with age between $a^{\prime }$ and $a^{\prime }+da^{\prime }$
infected when aged between $\tau $ and $\tau +d\tau $ and treated
continuously after $l$, that is, the treatment is initiated between $l$ and $%
l+dl.$ Therefore, the fraction of the unprotected sexual contacts given by
equation (\ref{0}), with individuals with age $a^{\prime }$ and $a^{\prime
}+da^{\prime }$ infected when aged between $\tau $ and $\tau +d\tau $ and
treated between $l$ and $l+dl$ is 
\begin{equation}
\beta _{0}(a,a^{\prime })da^{\prime }\frac{Y_{2}(a^{\prime },\tau
,l)da^{\prime }d\tau dl}{N(a^{\prime })da^{\prime }}X(a)da  \label{0c}
\end{equation}

Note that we assumed that the sexual behavior of susceptible towards treated
and untreated infected individuals is the same.

Let $g_{1}(a^{\prime }-l,l-\tau )$ be the probability of a susceptible
individual to get infected when making an unprotected sexual contact with a
treated individual $Y_{2}(a^{\prime },\tau ,l)da^{\prime }d\tau dl$. This
depends on viral load which in turn depends on the time since infection $%
(a^{\prime }-\tau )$ and the time since the start of the treatment, and its
form is given by equation (\ref{21}). Therefore, the number of new
infections per unit time, due to treated individuals $Y_{2}(a^{\prime },\tau
,l)da^{\prime }d\tau dl$, is given by

\begin{equation}
\beta _{0}(a,a^{\prime })g_{1}(a^{\prime }-l,l-\tau )\frac{Y_{2}(a^{\prime
},\tau ,l)}{N(a^{\prime })}da^{\prime }d\tau dlX(a)da  \label{0d}
\end{equation}

After integrating equation (\ref{0b}) from $0$ to $a^{\prime }$ with respect
to $\tau $, and from $0$ to infinite with respect to $a^{\prime }$, and
integrating (\ref{0d}) from $\tau $ to $a^{\prime }$ with respect to $l$,
from $0$ to $a^{\prime }$ with respect to $\tau $, and from $0$ to infinity,
with respect to $a^{\prime }$, and summing the two contributions, we get the
so-called \textit{per capita force of infection,} $\lambda (a),$ defined as

\begin{equation}
\begin{array}{lll}
\lambda (a) & = & \int_{0}^{\infty }\int_{0}^{a^{\prime }}\beta
_{0}(a,a^{\prime })g(a^{\prime }-\tau )\frac{Y_{1}(a^{\prime },\tau )}{%
N(a^{\prime })}d\tau da^{\prime } \\ 
&  &  \\ 
& + & \int_{0}^{\infty }\int_{0}^{a^{\prime }}\int_{\tau }^{a^{\prime
}}\beta _{0}(a,a^{\prime })g_{1}(a^{\prime }-l,l-\tau )\frac{Y_{2}(a^{\prime
},\tau ,l)}{N(a^{\prime })}dld\tau da^{\prime }
\end{array}
\label{2}
\end{equation}

As we showed in a previous paper \cite{amaku2003}, the contact function $%
\beta _{0}(a,a^{\prime })$ must satisfy a symmetry relation. To see this,
let 
\begin{equation}
\beta _{0}(a,a^{\prime })da^{\prime }N(a)da  \label{2aa}
\end{equation}
be the number of unprotected sexual contacts individuals with age between $a$%
\ and $a+da$\ make with\ all individuals aged between $a^{\prime }$\ and $%
a^{\prime }+da^{\prime }$. This number should be equal to the number of
unprotected sexual contacts that individuals aged between $a^{\prime }$\ and 
$a^{\prime }+da^{\prime }$\ make with \ all individuals with age between $a$%
\ and $a+da$,

\begin{equation}
\beta _{0}(a^{\prime },a)daN(a^{\prime })da^{\prime }  \label{2ab}
\end{equation}
that is,

\begin{equation}
\frac{\beta _{0}(a,a^{\prime })}{N(a^{\prime })}=\frac{\beta _{0}(a^{\prime
},a)}{N(a)}  \label{2a}
\end{equation}

This can be satisfied if $\beta _{0}(a,a^{\prime })$ is of the form 
\begin{equation}
\beta _{0}(a,a^{\prime })=f(a,a^{\prime })\frac{N(a^{\prime })}{N}
\label{2b}
\end{equation}
where $N=$ $\int_{0}^{\infty }N(a)da$ is the total population and $%
f(a,a^{\prime })$ is a symmetric function of $a$ and $a^{\prime }$,
describing the per capita rate of unprotected sexual contacts(see equations 
\ref{12} and \ref{16}). Note that, due to the division by population size an
increase in the frequency of age groups entirely uninterested for a given
person will decrease his/hers sexual activity with those specific age
groups. However, this is reflected automatically in the form of $%
f(a,a^{\prime })$.

Substituting (\ref{2b}) in (\ref{2}) we get

\begin{eqnarray}
\lambda (a) &=&\frac{1}{N}\int_{0}^{\infty }\int_{0}^{a^{\prime
}}f(a,a^{\prime })g(a^{\prime }-\tau )Y_{1}(a^{\prime },\tau )d\tau
da^{\prime }  \nonumber \\
&&+\frac{1}{N}\int_{0}^{\infty }\int_{0}^{a^{\prime }}\int_{\tau
}^{a^{\prime }}f(a,a^{\prime })g_{1}(a^{\prime }-l,l-\tau )  \nonumber \\
&&Y_{2}(a^{\prime },\tau ,l)dld\tau da^{\prime }  \label{2c}
\end{eqnarray}

Now, the equation for $X(a)$ is 
\begin{equation}
\frac{dX(a)}{da}=-\lambda (a)X(a)-\mu X(a)  \label{3}
\end{equation}
where $\mu $ is the natural mortality rate of humans.

Equation (\ref{3}) can be integrated, resulting in

\begin{equation}
X(a)=X\left( 0\right) \exp \left[ -\int_{0}^{a}\lambda (s)ds-\mu a\right]
\quad .  \label{4}
\end{equation}

Let us define $h_{1}\left( a,\tau \right) $ as a function describing the
removal of individuals from the first infective condition by natural
mortality and additional mortality due to progression to AIDS (see equation 
\ref{15}), and $h_{2}\left( a,\tau \right) $ describing the removal by
antiretroviral treatment (see equation \ref{15a}). Then we can write:

\begin{equation}
Y_{1}(a,\tau )=Y_{1}(\tau ,\tau )h_{1}\left( a,\tau \right) h_{2}\left(
a,\tau \right)  \label{5}
\end{equation}
Now we have:

\begin{equation}
Y_{1}(a,a)=\lambda (a)X(a)  \label{6}
\end{equation}
Substituting equation (\ref{4}) in (\ref{6}), we have:

\begin{equation}
Y_{1}(a,a)=X(0)\lambda (a)\exp \left[ -\int_{0}^{a}\lambda (s)ds-\mu a\right]
\label{7}
\end{equation}

\bigskip Substituting (\ref{7}) in (\ref{5}) we get 
\begin{eqnarray}
Y_{1}(a,\tau ) &=&X(0)\lambda (\tau )\exp \left[ -\int_{0}^{\tau }\lambda
(s)ds-\mu \tau \right]  \nonumber \\
&&h_{1}\left( a,\tau \right) h_{2}\left( a,\tau \right)  \label{7b}
\end{eqnarray}

Substituting equation (\ref{7b}) in (\ref{2c}) we get: 
\begin{eqnarray}
\lambda (a) &=&\frac{X(0)}{N}\int_{0}^{\infty }\int_{0}^{a^{\prime
}}f(a,a^{\prime })g(a^{\prime }-\tau )  \nonumber \\
&&\lambda (\tau )e^{\left[ -\int_{0}^{\tau }\lambda (s)ds-\mu \tau \right]
}h_{1}\left( a^{\prime },\tau \right) h_{2}\left( a^{\prime },\tau \right)
d\tau da^{\prime }  \nonumber \\
&&+\frac{1}{N}\int_{0}^{\infty }\int_{0}^{a^{\prime }}\int_{\tau
}^{a^{\prime }}f(a,a^{\prime })  \nonumber \\
&&g_{1}(a^{\prime }-l,l-\tau )Y_{2}(a^{\prime },\tau ,l)dld\tau da^{\prime }
\label{8}
\end{eqnarray}

Assuming a antiretroviral treatment rate $\nu \left( \tau ,a\right) $, we
have 
\begin{equation}
Y_{2}(l,\tau ,l)=Y_{1}(l,\tau )\nu \left( \tau ,l\right)  \label{9}
\end{equation}
and 
\begin{equation}
Y_{2}(a,\tau ,l)=Y_{2}(l,\tau ,l)h_{3}\left( a,\tau ,l\right)  \label{9aa}
\end{equation}
where $h_{3}\left( a,\tau ,l\right) $ is a function describing the removal
of individuals from the treated condition by mortality (see equation \ref{25}
below). Then, substituting equations (\ref{7b}) in (\ref{9}) and the
resulting equation in equation (\ref{9aa}), we get 
\begin{eqnarray}
Y_{2}(a,\tau ,l) &=&X(0)\lambda (\tau )\exp \left[ -\int_{0}^{\tau }\lambda
(s)ds-\mu \tau \right]  \nonumber \\
&&h_{1}\left( l,\tau \right) h_{2}\left( l,\tau \right) h_{3}\left( a,\tau
,l\right) \nu \left( \tau ,l\right)  \label{9a}
\end{eqnarray}
so that we have 
\begin{eqnarray}
\lambda (a) &=&\frac{X(0)}{N}\int_{0}^{\infty }\int_{0}^{a^{\prime }}\lambda
(\tau )e^{-\int_{0}^{\tau }\lambda (s)ds-\mu \tau }  \nonumber \\
&&f(a,a^{\prime })g(a^{\prime }-\tau )h_{1}\left( a^{\prime },\tau \right)
h_{2}\left( a^{\prime },\tau \right) d\tau da^{\prime }  \nonumber \\
&&+\frac{X(0)}{N}\int_{0}^{\infty }\int_{0}^{a^{\prime }}\int_{\tau
}^{a^{\prime }}\nu \left( \tau ,l\right) h_{1}\left( l,\tau \right)
h_{2}\left( l,\tau \right) h_{3}\left( a^{\prime },\tau ,l\right)  \nonumber
\\
&&f(a,a^{\prime })g_{1}(a^{\prime }-l,l-\tau )dld\tau da^{\prime }
\label{10}
\end{eqnarray}

Equation (\ref{10}) always has $\lambda (a)=0$ as a solution. Depending on
the functions $f(a,a^{\prime })g(a^{\prime }-\tau )$, $f(a,a^{\prime
})g_{1}(a^{\prime }-l,l-\tau )$ and on the parameters of $f(a,a^{\prime
})g(a^{\prime }-\tau )$, $f(a,a^{\prime })g_{1}(a^{\prime }-l,l-\tau )$, $%
h_{1}\left( a^{\prime },\tau \right) $, $h_{2}\left( a^{\prime },\tau
\right) $, $h_{3}\left( a^{\prime },\tau ,l\right) $ and $\nu \left( \tau
,l\right) $, it may have another unique positive solution \cite{nos2001}.
The condition for equation (\ref{10}) to have another solution, that is $%
\lambda (a)\neq 0$ defines the threshold above which the infection can
establish itself in the population \cite{lopez}, \cite{nos93}.

If we assume that $X(0)$ is proportional to the total population, $X(0)=bN$,
where $b$ is a constant, equation (\ref{10}) becomes independent on the
population size. In this paper we take $b=\mu $, for simplicity.

\subsection{Parenterally transmitted HIV}

In this subsection we model the parenteral transmission of HIV due to
sharing contaminated syringes and needles or contaminated blood and blood
products.

We consider that sexual transmission among these individuals is negligible
when compared with the parenterally form of transmission. However, their
contribution to the sexual spread of the infection will be considered in
this section.

Let $Y_{1}^{I}(a^{\prime },\tau )da^{\prime }d\tau $ be the number of
individuals parenterally infected with HIV (PI), aged between $a^{\prime }$
and \thinspace $a^{\prime }+da^{\prime }$, who acquired the infection when
aged between $\tau $ and $\tau +d\tau .$ We consider that PIs enter this
group when aged $a_{1}$. We also consider that they form a proportion $\eta $
of the population. Calling $\xi $ the rate of sharing syringes (or receiving
blood or blood products) multiplied by the probability of getting the
infection if the syringe (or blood or blood products) is (are) infected, we
have: 
\begin{eqnarray}
Y_{1}^{I}(a^{\prime },\tau ) &=&X(0)\eta \xi \exp \left[ -\mu \tau \right]
\exp \left[ -\xi \left( \tau -a_{1}\right) \right]  \nonumber \\
&&\theta \left( \tau -a_{1}\right) h_{1}(a^{\prime },\tau )h_{2}(a^{\prime
},\tau )  \label{10a}
\end{eqnarray}
where $h_{1}(a^{\prime },\tau )$ and $h_{2}(a^{\prime },\tau )$ were defined
in the previous section and are the rates of removal from the infective
class by death and treatment, respectively. Note that we have assumed that
in the case of drug users, they remain drug users for the rest of their
lives and that their mortality rate is not affected by the drug addiction.
This simplification is partially supported by field work we carried out in
the past \cite{tatico}, when we demonstrated that the average time of drug
usage of the studied community was found to be around 10 years.

Assuming an antiretroviral treatment rate $\nu \left( \tau ,a\right) $, we
have 
\begin{equation}
Y_{2}^{I}(l,\tau ,l)=Y_{1}^{I}(l,\tau )\nu \left( \tau ,l\right)  \label{10b}
\end{equation}
and 
\[
Y_{2}^{I}(a,\tau ,l)=Y_{2}^{I}(l,\tau ,l)h_{3}\left( a,\tau ,l\right) 
\]
where $h_{3}\left( a,\tau ,l\right) $ is a function describing the removal
of individuals from the treated condition by mortality. Then, we get 
\begin{equation}
Y_{2}^{I}(a,\tau ,l)=h_{1}(a^{\prime },\tau )h_{2}(a^{\prime },\tau
)h_{3}\left( a,\tau ,l\right) \nu \left( \tau ,l\right)  \label{10c}
\end{equation}

Dividing equations (\ref{10a}) and (\ref{10c}) by $N$, and adding to the
corresponding terms in equation (\ref{2c}), we finally get, instead of (\ref
{7b}) and (\ref{9a}): 
\begin{eqnarray}
Y_{1}(a,\tau ) &=&X(0)\left( 1-\eta \right) \lambda (\tau )\exp \left[
-\int_{0}^{\tau }\lambda (s)ds-\mu \tau \right]  \nonumber \\
&&h_{1}\left( a,\tau \right) h_{2}\left( a,\tau \right) +\eta X(0)\xi \exp
\left[ -\mu \tau \right]  \nonumber \\
&&\exp \left[ -\xi \left( \tau -a_{1}\right) \right] \theta \left( \tau
-a_{1}\right) h_{1}\left( a,\tau \right) h_{2}\left( a,\tau \right)
\label{10c1}
\end{eqnarray}
and 
\begin{eqnarray}
Y_{2}(a,\tau ,l) &=&X(0)\left( 1-\eta \right) \lambda (\tau )\exp \left[
-\int_{0}^{\tau }\lambda (s)ds-\mu \tau \right]  \nonumber \\
&&h_{1}\left( l,\tau \right) h_{2}\left( l,\tau \right) h_{3}\left( a,\tau
,l\right) \nu \left( \tau ,l\right)  \nonumber \\
&&+\eta X(0)\xi \exp \left[ -\mu \tau \right] \exp \left[ -\xi \left( \tau
-a_{1}\right) \right]  \nonumber \\
&&\theta \left( \tau -a_{1}\right) h_{1}\left( l,\tau \right) h_{2}\left(
l,\tau \right) h_{3}\left( a,\tau ,l\right) \nu \left( \tau ,l\right)
\label{10d}
\end{eqnarray}

Finally, instead of (\ref{10}) we get 
\begin{eqnarray}
\lambda (a) &=&\frac{X(0)}{N}\int_{0}^{\infty }\int_{0}^{a^{\prime }}\left(
1-\eta \right) \lambda (\tau )  \nonumber \\
&&\exp \left[ -\int_{0}^{\tau }\lambda (s)ds-\mu \tau \right] f(a,a^{\prime
})g(a^{\prime }-\tau )h_{1}\left( a^{\prime },\tau \right)  \nonumber \\
&&h_{2}\left( a^{\prime },\tau \right) d\tau da^{\prime }+\frac{X(0)}{N}%
\int_{0}^{\infty }\int_{0}^{a^{\prime }}\eta \exp \left[ -\mu \tau \right] 
\nonumber \\
&&\exp \left[ -\xi \left( \tau -a_{1}\right) \right] \theta \left( \tau
-a_{1}\right) f(a,a^{\prime })g(a^{\prime }-\tau )  \nonumber \\
&&h_{1}\left( a^{\prime },\tau \right) h_{2}\left( a^{\prime },\tau \right)
d\tau da^{\prime }+\frac{X(0)}{N}\int_{0}^{\infty }\int_{0}^{a^{\prime
}}\left( 1-\eta \right)  \nonumber \\
&&\lambda (\tau )\exp \left[ -\int_{0}^{\tau }\lambda (s)ds-\mu \tau \right]
f(a,a^{\prime })g(a^{\prime }-\tau )  \nonumber \\
&&\int_{\tau }^{a^{\prime }}\nu \left( \tau ,l\right) h_{1}\left( l,\tau
\right) h_{2}\left( l,\tau \right) h_{3}\left( a^{\prime },\tau ,l\right)
f(a,a^{\prime })  \nonumber \\
&&g_{1}(a^{\prime }-l,l-\tau )dld\tau da^{\prime }+\frac{X(0)}{N}%
\int_{0}^{\infty }\int_{0}^{a^{\prime }}\eta  \nonumber \\
&&\exp \left[ -\mu \tau \right] \exp \left[ -\xi \left( \tau -a_{1}\right)
\right] \theta \left( \tau -a_{1}\right)  \nonumber \\
&&\int_{\tau }^{a^{\prime }}\nu \left( \tau ,l\right) h_{1}\left( l,\tau
\right) h_{2}\left( l,\tau \right) h_{3}\left( a^{\prime },\tau ,l\right) 
\nonumber \\
&&f(a,a^{\prime })g_{1}(a^{\prime }-l,l-\tau )dld\tau da^{\prime }
\label{10e}
\end{eqnarray}

As mentioned before, if we assume that $X(0)$ is proportional to the
population, for instance, $X(0)=\mu N$, the equation (\ref{10e}) becomes
independent of the population size.

\section{A schematic model}

In this section we propose forms for the different functions involved in
equation (\ref{10e}).

Let us begin with an untreated population, for which $\nu (\tau ,l)=0$ and $%
h_{2}(a^{\prime },\tau )=1$ so that the second part of the integral equation
(\ref{10e}) vanishes.

As in \cite{nos2001}, let us define the rate $f(a,a^{\prime }),$ in a very
schematic form:

\begin{equation}
f(a,a^{\prime })=f_{0}(a,a^{\prime })\theta (a-a_{0})\theta (a^{\prime
}-a_{0})  \label{12}
\end{equation}
where $f_{0}(a,a^{\prime })$ is the rate of unprotected sexual contacts and $%
a_{0}$ is the age at which individuals enter in the risk behavior group. The
Heaviside functions $\theta (a-a_{0})~$and $\theta (a^{\prime }-a_{0})$ mean
that sexual activity begins after age $a_{0}$.

For the function $f_{0}(a,a^{\prime })$, which describes the age preferences
in acquisition of new partners and the age decline in the sexual activity,
we propose: 
\begin{equation}
f_{0}(a,a^{\prime })=Q\beta _{3}(a)\beta _{3}(a^{\prime })\beta
_{4}(a-a^{\prime })  \label{16}
\end{equation}
where 
\begin{equation}
\beta _{3}(x)=\frac{1}{\sqrt{2\pi }\sigma _{1}}e^{-\frac{(x-M)^{2}}{\sigma
_{1}^{2}}}  \label{17}
\end{equation}
and 
\begin{equation}
\beta _{4}(a-a^{\prime })=\frac{1}{\sqrt{2\pi }\sigma _{2}}e^{-\frac{%
(a-a^{\prime })^{2}}{\sigma _{2}^{2}}}  \label{18}
\end{equation}
The parameter $M$ is the age of maximum sexual activity and the constant $Q$
is adjusted to give the assumed average number of unprotected sexual
contacts per unit time. Those forms for the $\beta $-functions (\ref{17})
and (\ref{18}) were chosen for convenience only and, although not supported
by social studies, they conform with the following facts: a) sexual activity
increases with age up to a maximum, decreasing thereafter; b) age
preferences of an individual are distributed around a maximum value which we
assumed to occur at the same age of the individual. In order to facilitate
the calculations we made both functions as symmetric around a central value
and we assumed a minimum age $a_{0},$ below which there is no sexual
activity. In addition we assumed that people without treatment, people
undergoing treatment, and even people with full blown AIDS have the same
sexual behavior. This assumption is supported by preliminary data in our
community of HIV patients (Bueno, personal communication).

Let $A^{NT}(a^{\prime }-\tau )$ be the viraemia level of non-treated
individuals which depends on the time interval since the infection. For $%
g(a^{\prime }-\tau )$\ we take the form 
\begin{equation}
g(a^{\prime }-\tau )=I(A^{NT}(a^{\prime }-\tau ))  \label{18a}
\end{equation}
where $I(A^{NT})$\ is a function representing the transmission of at least
one infective viral inocula, given the viremic level $A^{NT}.$\ As in \cite
{nos2001} we assume that 
\begin{equation}
I\left( A^{NT}\right) =(c^{\prime }+c^{\prime \prime }\log A^{NT})\theta
\left( \log A^{NT}-3\right)  \label{13b}
\end{equation}
where $c^{\prime }=-3.35\times 10^{-3}$ and $c^{\prime \prime }=1.2$ $\times
10^{-3}$are parameters obtained by fitting equation (\ref{13b}) to the data
by Gray et al. \cite{gray} (an almost identical relationship between log of
viral load and transmissibility was found by using data from Vella et al 
\cite{vella} and Garcia et al \cite{garcia}, and it is supported by the
observations of Fideli et al.\cite{fideli} and Quin et al. \cite{quin}). The
Heaviside function $\theta \left( \log A^{NT}-3\right) $ is such that if $%
\log A^{NT}\leq 3$ there is no transmission.

For the removal of infected individuals, we assumed that when $\left(
a^{\prime }-\tau \right) <$ $L_{c},$ where $L_{c}$ is a given critical
moment when individuals reach a certain viraemia level and are defined as
AIDS patients, dies with rate $\mu $. When $\left( a^{\prime }-\tau \right)
>L_{c}$, individuals are subjected to an additional, disease specific, death
rate, $\alpha $. A simple form for this removal function is:

\begin{eqnarray}
h_{1}\left( a^{\prime },\tau \right) &=&e^{-\mu \left( a^{\prime }-\tau
\right) }\theta \left( L_{c}-(a^{\prime }-\tau )\right)  \nonumber \\
&&+e^{-\mu \left( a^{\prime }-\tau \right) -\alpha \left( (a^{\prime }-\tau
)-L_{c}\right) }\theta \left( (a^{\prime }-\tau )-L_{c}\right)  \label{15}
\end{eqnarray}

Let us now consider the treated population. The variables and parameters
already defined for the untreated population remain the same. We must only
define values for $h_{2}(a^{\prime },\tau )$, $\nu (\tau ,l)$, $%
g_{1}(a^{\prime }-l,l-\tau )$ and $h_{3}(a^{\prime },\tau ,l)$, where $l$ is
the moment in the history of the infection at which individuals begin to be
treated.

Let $A^{T}(a^{\prime }-l,l-\tau )$ denote the viraemia level after
antiretroviral treatment. We assumed that the viraemia level after
antiretroviral treatment is reduced by a certain factor, $\Delta (l-\tau )$,
increasing thereafter as 
\begin{equation}
A^{T}(a^{\prime }-l,l-\tau )=\frac{A^{NT}(l-\tau )\Psi (a^{\prime }-l)}{%
\Delta (l-\tau )}  \label{15b}
\end{equation}
where $\Psi (a^{\prime }-l)$ is a given function with $\Psi (0)=1$. Note
that with such functions, the log of the viraemia level just after the
beginning of the antiretroviral treatment starts with $\log \left[
A^{NT}(l-\tau )\right] -\log \left[ \Delta (l-\tau )\right] $ and increases
with $\log \left[ \Psi (a^{\prime }-l)\right] $ thereafter.

Figure 3a schematically illustrates the natural history of HIV infection in
the absence of antiretroviral treatment. Figure 3b illustrates the effect of
antiretroviral treatment on the viraemia level.

\begin{center}
\textbf{Figure 3a}

\textbf{Figure 3b}
\end{center}

In Figure 3a the viraemia level reaches a maximum immediately after the
moment $\tau $ of the infection and drops rapidly after some few weeks,
probably due to the effect of the immune system. Viraemia then starts to
raise, and simultaneously a decrease in the counting of CD$_{4}$
T-lymphocytes is observed, until a critical viral load level is reached ($%
L_{c}$), when the immune system breaks down and full blown AIDS develops.
This period lasts for 10 - 15 years.

Figure 3b represents the assumed natural history of HIV infection in the
presence of antiretroviral treatment, which begins at the age of infection
between $l$ and $l+dl.$ The treatment starts at any moment after the viral
load reaches 30,000 copies/ml \cite{dstaids}(we are well aware, however,
that in real life other clinical indicators of antiretroviral treatment,
like CD$_{4}$ counting, are used to begin the antiretroviral treatment.). We
assumed that the treatment causes a sharp decrease in the viral load, and
that it loses its effect immediately so that the viral load starts to rise
log-linearly again. This intends to model the appearance of resistant
strains, which we assumed to have the same virulence as the original strain.
Therefore, the assumption of log-linearity in the viral load curve is
entirely hypothetical and was intended to mimic an immediate development of
full resistance by HIV to the treatment , which is the worst epidemiological
scenario. In this situation, we are assuming that replication of HIV is no
longer constrained by treatment. Another assumption of our model is that it
takes no account of the role of the immunity system on HIV replication. With
the above assumptions, the viral load after treatment increases until a
critical level is eventually reached ($L_{c}^{\prime }$), when the immune
system breaks down and full blown AIDS develops. Hence, the treatment makes
the period without AIDS longer than 10 years.

We assumed also that the antiretroviral treatment schedule is given by $\nu
(\tau ,l)=\nu \times \theta \left( (l-\tau )-a_{t}\right) $, and it begins,
as mentioned above, when viraemia reaches\emph{\ }30,000 copies/ml, which
occurs at the infection age $(l-\tau )=a_{t}$ ($\simeq 6.7$ years)$.$ Thus 
\begin{equation}
h_{2}(l,\tau )=e^{-\nu \times \left( (l-\tau )-a_{t}\right) \times \theta
\left( (l-\tau )-a_{t}\right) }\quad .  \label{15a}
\end{equation}

Since we assume that the treatment affects only the viral load, we may
define $g_{1}(a^{\prime }-l,l-\tau )$ as: 
\begin{equation}
g_{1}(a^{\prime }-l,l-\tau )=I\left( A^{T}(a^{\prime }-l,l-\tau )\right)
\label{21}
\end{equation}

Finally, for $h_{3}(a,\tau ,l)$ we may also assume a simple form, similar to 
$h_{1}\left( a^{\prime }-\tau \right) $: 
\begin{eqnarray}
h_{3}(a^{\prime },\tau ,l) &=&h_{3}(a^{\prime },l)=e^{-\mu \left( a^{\prime
}-l\right) }\theta \left( L_{c}^{\prime }-(a^{\prime }-l)\right)  \nonumber
\\
&&+e^{-\mu \left( a^{\prime }-l\right) -\delta \left( (a^{\prime
}-l)-L_{c}^{\prime }\right) }\theta \left( (a^{\prime }-l)-L_{c}^{\prime
}\right)  \label{25}
\end{eqnarray}
where $L_{c}^{\prime }$ is the new critical moment when treated individuals
reach a certain viraemia level and are defined again as AIDS patients. Those
patients are subjected to a new differential mortality rate $\delta $ due to
AIDS.

Within this framework we can obtain different particular models by choosing
suitable functions $A^{NT}$, $A^{T}$\emph{\ }and constants.

We adopted, for the non-treated patients, the following average values for $%
A_{i}$ and $L_{i}$: 
\begin{equation}
A^{NT}(a^{\prime }-\tau )=\left\{ 
\begin{array}{lll}
A_{1}(a^{\prime }-\tau ) & = & 10^{6} \\ 
& for & 0\leq (a^{\prime }-\tau )<L_{1} \\ 
A_{2}(a^{\prime }-\tau ) & = & 10^{3+0.22(a^{\prime }-\tau -L_{1})} \\ 
& for & L_{1}\leq (a^{\prime }-\tau )<L_{c} \\ 
A_{3}(a^{\prime }-\tau ) & = & 10^{6} \\ 
& for & (a^{\prime }-\tau )\geq L_{c}
\end{array}
\right.  \label{26}
\end{equation}
where $L_{1}=$ $6$ weeks and $L_{c}-L_{1}=\frac{3}{0.22}$ years is the
period of time it takes for the viraemia to reach $10^{6}\,$RNA copies per
ml since the beginning of the second phase.

Let us now model the effectiveness of antiretroviral treatment by an
instantaneous reduction on the viral load immediately after the introduction
of the treatment. We consider several effectiveness levels of antiretroviral
treatment, by varying the reduction of the viral load. When the reduction is
by a factor greater than $10^{2}$, the antiretroviral treatment is known as
HAART (Highly Active AntiRetroviral Treatment). If the reduction is lesser
than $10^{2}$, the antiretroviral treatment is known as non-HAART. Again,
for the sake of simplicity, we drop the superscript $T$ from the viral load $%
A$ and introduce a subscript $j$ $=4,5$, describing the two possible phases
of the natural history of the infection after treatment. Note that we
assumed no treatment during the first phase. 
\[
A^{T}((a^{\prime }-l),(l-\tau ))= 
\]
\begin{equation}
\left\{ 
\begin{array}{lll}
A_{4}((a^{\prime }-l),(l-\tau )) & = & 10^{3-K+0.22(l-\tau
-L_{1})+0.22(a^{\prime }-l)} \\ 
& for & 0\leq (a^{\prime }-l)<L_{c}^{\prime } \\ 
A_{5}((a^{\prime }-l),(l-\tau )) & = & 10^{6} \\ 
& for & (a^{\prime }-l)\geq L_{c}^{\prime }
\end{array}
\right.  \label{35}
\end{equation}
where $L_{c}^{\prime }=\frac{3+K-0.22(l-\tau -L_{1})}{0.22}$ years, so that $%
\Psi (a^{\prime }-l)=10^{0.22(a^{\prime }-l)}$ , similar to the non-treated
patients, and $\Delta (l-\tau )=K\geq 2$ for HAART treatment and $\Delta
(l-\tau )=K<2$ for non-HAART treatment. In the numerical simulations several
values of $K$ will be considered. Note that $L_{c}^{\prime }$ is the period
of time it takes for the viraemia of treated individuals to reach $10^{6}\,$%
RNA copies per ml since the beginning of antiretroviral treatment.

With the above models of viraemia, $g(a^{\prime }-\tau )$\ takes the form

\begin{equation}
g(a^{\prime }-\tau )=\left\{ 
\begin{array}{l}
I(A_{1}((a^{\prime }-\tau )))\,\quad for\quad 0\leq (a^{\prime }-\tau )<L_{1}
\\ 
I(A_{2}((a^{\prime }-\tau )))\,\quad for\quad L_{1}\leq (a^{\prime }-\tau
)<L_{c} \\ 
I(A_{3}((a^{\prime }-\tau )))\,\quad for\quad (a^{\prime }-\tau )\geq L_{c}
\end{array}
\right. ~,  \label{13}
\end{equation}
where $I(A_{1})$\ is a function defined by equation (\ref{13b}),
representing the probability of transmission of at least one infective viral
inocula. Correspondingly, $g_{1}(a^{\prime }-l,\tau )$\ takes the form

\begin{equation}
g_{1}(a^{\prime }-l,l-\tau )=\left\{ 
\begin{array}{l}
I(A_{4}((a^{\prime }-l),(l-\tau )))\, \\ 
\quad for\quad 0\leq (a^{\prime }-l)<L_{c}^{\prime } \\ 
I(A_{5}((a^{\prime }-l),(l-\tau ))) \\ 
\,\quad for\quad (a^{\prime }-l)\geq L_{c}^{\prime }
\end{array}
\right.  \label{22}
\end{equation}

Finally, for the parenterally transmitted branch of the infection we take $%
\eta =0.01$, $\xi =0.05$ per year and $a_{1}=15$ years \cite{tatico}, \cite
{carlini}.

\section{Numerical results}

In order to analyze the model's performance against an HIV endemic situation
we solved equation (\ref{10e}) numerically for several treatment schedules.
For this we divided the population into four classes. Each class obeys an
equation like equation (\ref{10e}) and intends to mimic a specific risk
group, namely, group I (GI), with very low level of sexual promiscuity (84\%
of the total population), group II\ (GII)\ with moderate levels of sexual
promiscuity\textit{\ }(10\%), group III\ (GIII) with high levels of sexual
promiscuity (5\%), and group IV (GIV) with very high levels of sexual
promiscuity (1\%). This somewhat arbitrary division is based on the actual
classes of risk recognizable in real populations. So GI could represent the 
\textit{general population }, GII \textit{promiscuous heterosexuals}, GIII 
\textit{male homosexuals}, and GIV \textit{commercial sex workers and their
clients}.We also assume that in each class there is a fraction $\eta $ of 
\textit{PIs}, who got infected by contaminated syringes/needles or
blood/blood products. With respect to sexual contacts, it should be stressed
that we are only interested in unprotected sexual contacts with new
partners. In the functions $\beta _{3}$\ and $\beta _{4}$, which describe
the sexual behavior, we set, for all classes, $\sigma _{1}=10$\ years, $%
\sigma _{2}=15$\ years, $M=25$\ years. The mortality (and fertility) rate
was taken $\mu =1/70\ years^{-1}$. The initial age of sexual life $a_{0}$\
was set to 15 years.\emph{\ }

The parameter $Q$ is related, in the absence of the infection, to the per
capita number of unprotected sexual contacts. We can, then, calculate the
per capita number of unprotected sexual contacts in the population, $\Phi $,
as 
\begin{eqnarray}
\Phi &=&\frac{1}{N}\int_{0}^{\infty }\int_{0}^{\infty }\beta (a,a^{\prime
})N(a)da^{\prime }da  \nonumber \\
&=&\mu \int_{0}^{\infty }\int_{0}^{\infty }\frac{f_{0}(a,a^{\prime })}{N}%
\theta (a-a_{0})\theta (a^{\prime }-a_{0})N(a^{\prime })e^{-\mu a}da^{\prime
}da  \label{30}
\end{eqnarray}
using equation (\ref{12}). Using equation (\ref{16}) we get the relation
between the per capita number of unprotected sexual contacts in the absence
of infection, $\Phi $, and $Q$: 
\begin{equation}
\Phi =\mu ^{2}Q\int_{a_{0}}^{\infty }\int_{a_{0}}^{\infty }\beta
_{3}(a)\beta _{3}(a^{\prime })\beta _{4}(a-a^{\prime })e^{-\mu a^{\prime
}}e^{-\mu a}da^{\prime }da  \label{31}
\end{equation}

The parameter $Q$ was adjusted to give the estimated average number of
unprotected sexual contacts per unit time, $\Phi $. The results, compatible
with the literature, for each class were $Q=6\times 10^{6}$ (GI), $%
Q=1.38\times 10^{7}$ (GII), $Q=1.45\times 10^{7}$ (GIII), $Q=1.55\times
10^{7}$ (GIV), as can be seen in table 1.

With respect to the PI arm of transmission, as mentioned above, we chose the
parameters $a_{1}=15$ years, $\eta =0.001$ and $\xi =0.05/year.$

Now, in the presence of HIV infection, without any treatment, the incidence
of sexually transmitted HIV $(i)$ is defined as the number of new cases of
HIV infection per year per person and was calculated as

\begin{equation}
i=\frac{\left( 1-\eta \right) \left[ \int_{0}^{\infty }\lambda
(a)X(a)da\right] }{N}  \label{32}
\end{equation}
and the incidence of parenterally transmitted HIV $(i^{I})$ is also defined
as the number of new cases of HIV infection per year per person and was
calculated as 
\begin{equation}
i^{I}=\eta \mu \xi \int_{0}^{\infty }\exp \left[ -\mu a\right] \exp \left[
-\xi (a-a_{1}\right] \theta (a-a_{1})da  \label{32a}
\end{equation}
With the adopted parameters $a_{1}=15$ years, $\eta =0.001$ and $\xi =0.05$
per year, we have $i^{I}=9\times 10^{-6}$ per person-year.

The prevalence of sexually transmitted HIV $(p)$ in the absence of treatment
is given by 
\begin{equation}
p=\int_{0}^{\infty }\int_{0}^{a}y_{1}(a,\tau )d\tau da  \label{33}
\end{equation}
where

\begin{equation}
y_{1}(a,\tau )=\frac{Y_{1}(a,\tau )}{N}  \label{33'}
\end{equation}
where $Y_{1}(a,\tau )$ is given by equation (\ref{10c1}) and $\lambda (a)$
given by equation (\ref{10e}), with $\nu (\tau ,l)=0$ and $h_{2}(a^{\prime
},\tau )=1$. The prevalence of parenterally transmitted HIV $(p^{I})$,
without treatment, is given by 
\begin{eqnarray}
p^{I} &=&\int_{0}^{\infty }\int_{0}^{a}\eta \mu \xi \exp \left[ -\mu \tau
\right]  \nonumber \\
&&\exp \left[ -\xi \left( \tau -a_{1}\right) \right] \theta \left( \tau
-a_{1}\right) h_{1}(a,\tau )d\tau da.  \label{33a}
\end{eqnarray}
With the adopted parameters $a_{1}=15$ years, $\eta =0.001$ and $\xi
=0.05/year$, we have $p^{I}=5.9\times 10^{-4}$.

The results of the simulation for the per capita number of unprotected
sexual contacts, the prevalence and the incidence of HIV infection at
equilibrium are given in table 1:

\begin{center}
\begin{tabular}{cccccc}
\hline
\multicolumn{6}{c}{\textbf{Table 1: }prevalence and incidence of HIV
infection at equilibrium} \\ \hline
&  & without PI &  & with PI &  \\ \hline
class(\%) & $\Phi (years^{-1})$ & incidence & prevalence & incidence & 
prevalence \\ \hline
GI(84\%) & 6.60 \cite{oxman} & 0.00 & 0.00 & 8.51E-05 & 1.25E-03 \\ 
GII(84\%) & 15.30 \cite{recrutas} & 7.09E-04 & 1.08E-02 & 1.20E-03 & 1.89E-02
\\ 
GIII(84\%) & 16.00 \cite{craib} & 2.60E-03 & 4.50E-02 & 2.75E-03 & 4.77E-02
\\ 
GIV(84\%) & 17.10 \cite{sz} & 4.60E-03 & 9.00E-02 & 4.63E-03 & 9.17E-02 \\ 
\hline
Average(84\%) & 13.20 & 2.47E-04 & 4.23E-03 & 3.75E-04 & 6.24E-03 \\ \hline
\end{tabular}
\end{center}

The scenario presented in table 1 is not intended to fit real data but
rather to present qualitatively the effects of treatment. It is interesting,
however, that the prevalence results are compatible with regions with
intermediate levels of HIV transmission. By intermediate levels of HIV
transmission, we mean regions of the world with HIV/AIDS prevalences ranging
from 0.5\% to 1.2\%\cite{who}. Note that without the PI class there would be
no disease in the general population and the infection would be restricted
to the other three classes. However, as mentioned before, our model is too
schematic to guarantee that the above conclusion can be extended to any real
population.

Note also that in table 1 the ratio between $i$ and $\Phi $, which gives the
number of new cases of HIV infection per unprotected sexual relation with
each new partner. So, for instance, considering the class GIV, our results
point to a risk of acquiring HIV infection of approximately 3.0\% per
unprotected sexual contact. This result is of the same order of that
reported in the literature (for instance, in Thailand this risk was
estimated to be between 3.1\% and 5.6\% \cite{SW-Thai}).

In order to evaluate the impact of antiretroviral treatment on the incidence
and prevalence of HIV infection we simulated the model for values of $K$
(effectiveness of treatment) ranging from $0$ to $3$ and $\nu $ (intensity
of treatment) ranging from $0.05$ to $0.5$. The impact of antiretroviral
treatment on the incidence and prevalence of HIV infection in each class is
shown in figures 4 to 8. Figure 4 shows the effect of antiretroviral
treatment on the class representing GI ($Q=115,000$). In this class the risk
of contracting HIV is very low, reaching a prevalence without antiretroviral
treatment of the order of 0.1\%. In fact, this low prevalence is due to the
PI individuals, without whom the disease among the general population would
disappear, meaning that for $Q=115,000$ the infection is below the threshold
for its maintenance in the general population. Figures 4a and 4b show the
incidence and prevalence, respectively, as functions of the intensity of
antiretroviral treatment $\nu $ for six levels of effectiveness of
antiretroviral treatment $K$. Note that the infection drops monotonically
with $\nu $\ for all values of $K$.\emph{\ }

\begin{center}
\textbf{Figure 4a}

\textbf{Figure 4b}
\end{center}

Figure 5 shows the effect of antiretroviral treatment on the class
representing the subpopulation GII ($Q=138,000$). In this class the risk of
contracting HIV is higher than in the GI, described above. Without
antiretroviral treatment the prevalence of HIV infection reaches the order
of 2.0\%. Figures 5a and 5b show the incidence and prevalence, respectively,
as functions of the intensity of antiretroviral treatment $\nu $ for six
levels of effectiveness of antiretroviral treatment $K$. Note that for $%
K=0.5 $ the treatment results in higher prevalence and incidence than in its
absence, for all values of $\nu $. Moreover, both the incidence and the
prevalence show an initial increase with the treatment intensity and a
decrease after around $\upsilon =0.15$.

\begin{center}
\textbf{Figure 5a}

\textbf{Figure 5b}
\end{center}

To understand the phenomenon described above, one should consider the
following: first, a low intensity treatment means that the individuals, on
the average, start to be treated later than with a high intensity treatment;
secondly, starting the treatment implies in reducing the probability of HIV
transmission proportionally to the reduction in the log of viral load due to
the effect of the treatment; finally, treating individuals implies in a
longer survival period, therefore increasing the total number of sexual
contacts of those individuals. Hence, if the effectiveness of the treatment (%
$K$) does not reduce the log of the viral load sufficiently to decrease the
probability of transmission per sexual contact such as to compensate the
increased transmission of HIV due to the higher number of sexual contacts,
the total contribution of those individuals to HIV\ transmission will
increase. Otherwise it will decrease. This effect repeats itself for the
other simulations described below.

Figure 6 shows the effect of antiretroviral treatment on the class
representing the subpopulation of GIII individuals ($Q=145,000$). In this
class the risk of contracting HIV is higher than that of GII, described
above. Without antiretroviral treatment the prevalence of HIV infection
reaches the order of 4.8\%. Figures 6a and 6b show the incidence and
prevalence, respectively, as functions of the intensity of antiretroviral
treatment $\nu $ for six levels of effectiveness of antiretroviral treatment 
$K$. Note that, for $K=0.5$\ the prevalence and the incidence is greater
than without treatment for all values of $\nu .$ For $K\geq 1$, the
incidence drops monotonically with $\nu .$ However, for $K=1$, the
prevalence is greater than that without treatment for values of $\nu $ up to
approximately 0.8.

\begin{center}
\textbf{Figure 6a}

\textbf{Figure 6b}
\end{center}

Figure 7 shows the effect of antiretroviral treatment on the class
representing the subpopulation GIV ($Q=155,000$). In this class the risk of
contracting HIV is the highest among all classes considered. Without
antiretroviral treatment the prevalence of HIV infection reaches the order
of 9\%. Figures 7a and 7b show the incidence and prevalence, respectively,
as functions of the intensity of antiretroviral treatment $\nu $ for six
levels of effectiveness of antiretroviral treatment $K$.\emph{\ }Note that
both the incidence and the prevalence for $K=0.5$ are higher than those
without treatment, for all values of $\nu $. For $K=1$ and $K=1.5$, the
incidence drops monotonically but the prevalence is greater than that
without treatment for small values of $\nu $.

\begin{center}
\textbf{Figure 7a}

\textbf{Figure 7b}
\end{center}

Figure 8 shows the weighted average of the incidence (8a) and prevalence
(8b) curves over the entire population.\emph{\ }Note that both the incidence
and the prevalence for $K=0.5$ are higher than those without treatment, for
all values of $\nu $. For $K\geq 1$, both the incidence and prevalence drop
monotonically.

\begin{center}
\textbf{Figure 8a}

\textbf{Figure 8b}
\end{center}

\section{Comments and conclusions}

In this paper we presented a very simple model of the steady-state effect of
HAART on HIV incidence and prevalence. The model mimics current treatment
guidelines applied in Brazil. However, the model does not intend to fit the
data with any acceptable degree of accuracy since detailed information
necessary are not available. So, this paper intends to provide a conceptual
and mechanistic understanding of the possible long term effects of treatment
on the dynamics of HIV transmission. As mentioned by Anderson \cite{and},
one of the purposes of modelling is to help identifying areas in which
better epidemiological data is required to refine prediction and improve
understanding, guiding, in a way, field research. We hope our model can be
useful in pointing which parameters should be better determined in future
studies.

The model could be extended to allow the calculation of temporal evolution
of the effects of treatment on HIV incidence, as we did for rubella
vaccination in \cite{amaku2003}. At the present stage we calculated only
steady-state effects, which is important enough for assessing long term
effects of treatment on trends of HIV dynamics.

One important aspect of the model is that the population was divided into
four compartments according to their sexual activities. We also assumed
another compartment of PIs and we assumed that only they interact with the
other four different model compartments. The PIs were singled out because
they acquire the infection by a different route. It would appear natural to
consider that the risk groups would interact with each other. However, this
would introduce more unknown parameters and, therefore, for simplicity we
considered such an interaction as negligible. Note that this implies that if
an individual from any single risk group, for instance GI, has an
unprotected sexual contact with any other individuals from another group
he/she and his/her stable partner would be considered as belonging to this
class.

Our main conclusions are as follows:

\begin{enumerate}
\item  The disease, according to the model, is almost completely wiped out
when we consider the most effective treatment ($K=3.0$) simulated. This
conclusion should be taken with great care since it may be the result of
some 'mathematical pathology' taken to its extreme. In fact, the model
predicts that under this treatment regime the disease is maintained in the
population (all four classes) due to the interaction of individuals from the
other classes with PIs. Again, this conclusion should be taken with caution
since, by allowing strong interactions between the distinct sexual behavior
classes the disease might not disappear under treatment. In addition, note
that our simulated treatment represents the worst scenario (the immediate
evolution of complete resistance with no further alteration in the treatment
scheme), while in clinical practice modifications of treatment schemes
should always follows a significant increase in viraemia.

\item  The impact of the treatment on HIV incidence or prevalence depends on
the level of sexual activity of the subpopulations considered, being more
pronounced on the subpopulations with the highest sexual activity levels. By
impact of treatment we mean the difference between the pre-treatment level
of incidence or prevalence and the equilibrium attained with the maximum
intensity of treatment, $\nu $. This conclusion is valid only for the most
effective treatment scheme, $K=3.0$

\item  Inefficient treatment, $K<1.0$, can be prejudicial on the
subpopulations with high levels of sexual activity. For instance, in
populations with intermediate levels of sexual activity ($\Phi =0.153$ years$%
^{-1}$), the effect of inefficient treatment depends on the intensity of
treatment, $\nu ,$ in a curious way. For $\nu $ between 0 and 0.15, there is
an increase in both incidence and prevalence, which then drops thereafter.
The reason for this behavior is explained in the main body of the text.
\end{enumerate}

As a general comment, there are many ways to express the intensity of
transmission of an infection, among which the classical basic reproduction
ratio, $R_{0}$, the force of infection, $\lambda (a)$, the incidence and the
prevalence. The basic reproduction ratio is the greatest eigenvalue of the
Frechet derivative with respect to $\lambda (a)$ of the operator, which is
the right hand side of equation \ref{10e}, calculated at $\lambda (a)=0$
(see \cite{lopez}, \cite{frailty}). In this paper we calculated the
incidence and the prevalence of HIV as these are the parameters most used by
public health authorities to monitor HIV\ epidemic.

Finally, since HIV treatment begins late in the second phase (in practice
and in this model), the intensity and diversity of effects that result from
different treatment strategies is another evidence of the importance of the
asymptomatic phase of HIV infection on the spread of the virus. This has
already been pointed out \cite{letter}, contrasting with former opinions of
some infectious disease practitioners and epidemiologists.

\section{Acknowledgements}

This work was supported by grants from LIM01/HCFMUSP, CNPq, PRONEX\ and
FAPESP.

\newpage

\begin{center}
\textbf{Captions for the figures}
\end{center}

\textbf{Figure 1:} Reduction in the mortality by AIDS in the period between
1996 and 2001. The total number of averted deaths summed up to 90,000
patients in this period.

\textbf{Figure 2:} Reduction in the incidence rates in Brazil two years
after the introduction of universal highly active antiretroviral treatment.

\textbf{Figure 3:} (a) Model assumed to describe the natural variation of
HIV viraemia along the natural history of the infection (shaded area). The
infection happens at age between $\tau $ and $\tau $ +$d\tau $ and $L_{c}$
marks the beginning of full blown AIDS. (b) Model assumed to describe the
natural variation of HIV viraemia along the natural history of the infection
in presence of antiretroviral treatment (shaded area). The infection happens
at age between $\tau $ and $\tau $ +$d\tau $, the treatment begins at age
between$l$ and $l+dl$ and $L_{c}^{\prime }$ marks the beginning of full
blown AIDS.

\textbf{Figure 4:} (a) The incidence of HIV in the GI (general population)
group as a function of the treatment intensity rate $\nu $ for several
levels of effectiveness of antiretroviral treatment $K$. (b) The prevalence
of HIV in the GI (general population) group as a function of the treatment
intensity rate $\nu $ for several levels of effectiveness of antiretroviral
treatment $K$.

\textbf{Figure 5:} (a) The incidence of HIV in the class representing the
sub-population of GII (promiscuous heterosexuals) group, as a function of
the treatment intensity rate $\nu $ for several levels of effectiveness of
antiretroviral treatment $K$. (b) The prevalence of HIV in the class
representing the sub-population GII (promiscuous heterosexuals), as a
function of the treatment intensity rate $\nu $ for several levels of
effectiveness of antiretroviral treatment $K$.

\textbf{Figure 6:} (a) The incidence of HIV in the class representing the
sub-population GIII (male homosexuals), as a function of the treatment
intensity rate $\nu $ for several levels of effectiveness of antiretroviral
treatment $K$. (b) The prevalence of HIV in the class representing the
sub-population GIII\ (male homosexuals), as a function of the treatment
intensity rate $\nu $ for several levels of effectiveness of antiretroviral
treatment $K.$

\textbf{Figure 7: }(a) The incidence of HIV in the class representing the
sub-population GIV (sex workers) and their clients, as a function of the
treatment intensity rate $\nu $ for several levels of effectiveness of
antiretroviral treatment $K$. (b) The prevalence of HIV in the class
representing the sub-population GIV (sex workers) and their clients, as a
function of the treatment intensity rate $\nu $ for several levels of
effectiveness of antiretroviral treatment $K.$

\textbf{Figure 8:}(a) The incidence of HIV in the population average as a
function of the treatment intensity rate $\nu $ for several levels of
effectiveness of antiretroviral treatment $K$. (b) The prevalence of HIV in
the population average as a function of the treatment intensity rate $\nu $
for several levels of effectiveness of antiretroviral treatment $K.$

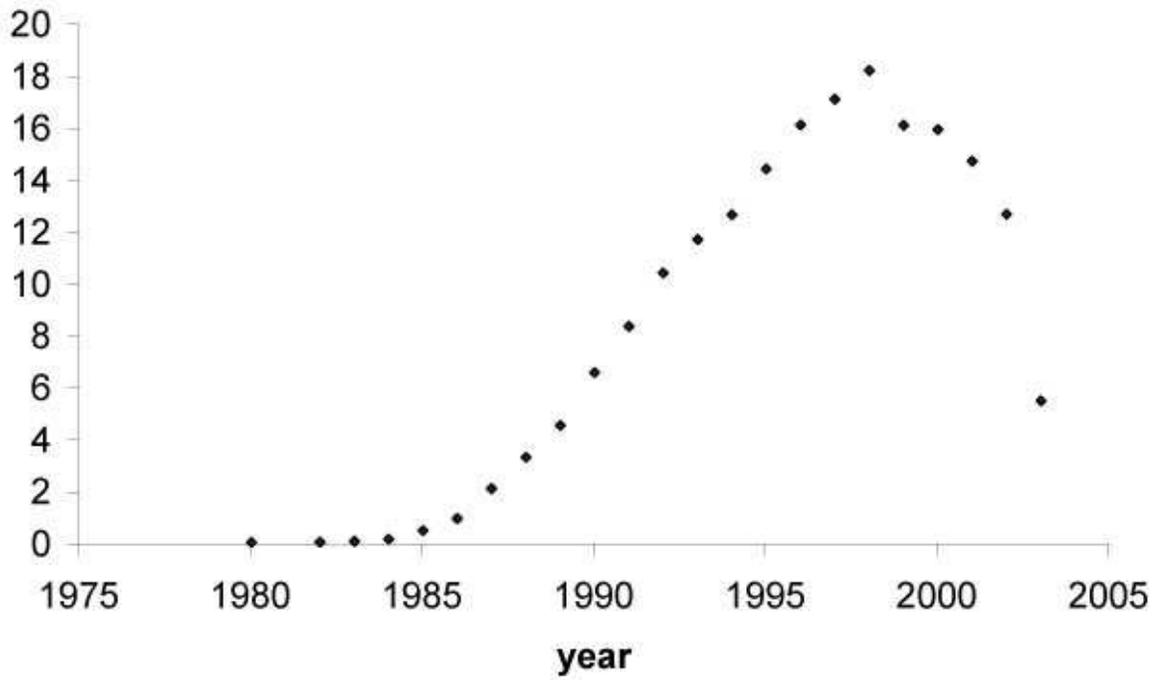
\end{document}